\documentclass[prb,preprint,showpacs]{revtex4}
\usepackage{graphicx}
\usepackage{amsmath,bm}
\usepackage{pstricks}

\begin{document}
\title{Functionalized adamantane: fundamental building blocks for
nanostructure self-assembly}

\author{J. C. Garcia$^{1}$, J. F. Justo$^{2}$, W. V. M. Machado$^{1}$,
and L. V. C. Assali$^{1}$}

\affiliation{$^1$ Instituto de F\'{\i}sica, Universidade de S\~ao Paulo,\\
CP 66318, CEP 05315-970, S\~ao Paulo, SP, Brazil \\
$^2$ Escola Polit\'ecnica, Universidade de S\~ao Paulo, \\
CP 61548, CEP 05424-970, S\~ao Paulo, SP, Brazil}

\date{August 12, 2009}

\begin{abstract}
We report first principles calculations on the electronic and structural
properties of chemically functionalized adamantane molecules, either in
isolated or crystalline forms. Boron and nitrogen functionalized molecules,
aza-, tetra-aza-, bora-, and tetra-bora-adamantane, were found to be very
stable in terms of energetics, consistent with available experimental
data. Additionally, a hypothetical molecular crystal in a zincblende
structure, involving the pair tetra-bora-adamantane and tetra-aza-adamantane,
was investigated. This molecular crystal presented a direct and large
electronic bandgap and a bulk modulus of 20 GPa. The viability of using
those functionalized molecules as fundamental building blocks for
nanostructure self-assembly is discussed.
\end{abstract}

\maketitle

\section{Introduction}
\label{sec1}

Carbon is unique in nature because it carries several competing hybridization
states for its valence electrons, leading to a number of stable, and sometimes
exotic, organized structures. Diamond and graphite crystalline structures,
with carbon respectively in sp$^3$ and sp$^2$ hybridizations, already present
remarkable properties, leading to applications ranging from cutting tools to
electronic devices. On the other hand, nanostructured forms of carbon,
which have been discovered over the last two decades, could potentially lead
to a much wider range of applications in a near future \cite{avouris}. Those
nanostructures, that included fullerenes, nanotubes, graphene, nanoribbons,
and others, have been the focus of intensive investigation. Fullerenes have
been considered as lubricants, nanotubes have been used to build nanodiodes
and nanotransistors \cite{tans}, and nanoribbons have been considered for
molecular electronics \cite{dai}.

Striking developments have been recently attained in building, separating,
manipulating, and functionalizing a different class of nanostructured carbon,
called diamondoids. They can be described as
molecular diamond, with carbon atoms in a diamond-like structure saturated
by hydrogen atoms in the surface. Due to the prevailing covalent nature of the
interatomic interactions, those molecules carry outstanding stability and
rigidity. Adamantane is the smallest diamondoid, with a C$_{10}$H$_{16}$
configuration, which consists of a single diamond-like carbon cage.
Larger diamondoids, such as diamantane, triamantane, and higher
polymantanes, are obtained by incorporating additional carbon cages to
adamantane.

Diamondoids have been known for almost 100 years, since their discovery in
petroleum. The recent identification and separation of higher
diamondoids \cite{dahl} allowed to envision several applications, such
as electron emission devices \cite{drumm,wlyang}, chemical sensors
\cite{basu}, biomarkers \cite{zdwang}, and pharmaceuticals \cite{schnell}.
They have also been considered as building blocks (BBs) to
build complex ordered nano-elements with sub-nanometric precision
\cite{mbb,mcintosh}. Using BBs represents a potentially competitive
building procedure for industrial applications within a bottom-up approach.
BBs can be thought as nanobricks, but manipulating them represents a
major challenge. Since adamantane is the smallest diamondoid,
it could be considered as a fundamental building block (FBB).
Current methods that allow positional control, such as
atomic force microscopy, would be unrealistically too slow to build
relevant systems in large scale. Therefore, new building procedures,
such as those based on self-assembly, should be considered.

Diamondoids carry several properties which make them suitable as BBs:
they have many sizes and forms, leading to several nanobrick types,
and their superior stability and rigidity benefits building robust
complex systems. However, diamondoids are fully hybridized with hydrogen,
leading to weak intermolecular interactions, and consequently brittle
crystalline structures. Functionalization could solve several of those
limitations, since it generates chemically active sites, leading to stronger
intermolecular interactions and stiff nanostructures. Such
functionalization could also help self-assembly, driving the system to
pre-determined organized configurations.

This investigation reports the physical properties of functionalized
adamantane molecules with boron and nitrogen to
form bora-, tetra-bora-, aza-, and tetra-aza-adamantane,
using first principles calculations. We found that
functionalization is strongly favorable and leads to at least two types
of FBBs: tetra-bora-adamantane and tetra-aza-adamantane. In such context,
we investigated a hypothetical crystalline structure, formed by a
combination of those two functionalized adamantane molecules. The paper is
organized as follow: section \ref{sec2} presents the theoretical model,
section \ref{sec3} presents the properties of pure and functionalized
adamantane, and section \ref{sec4} presents the properties of  a
crystalline structure formed by functionalized adamantane molecules.

\section{Methodology}
\label{sec2}

Theoretical modeling is an important tool to decipher the physical
properties of nanostructured systems. Diamondoids have been recently
investigated by first principles methodology: trends in stability and
electronic properties \cite{mcintosh,lu,yywang}, electron affinity
\cite{drumm}, trends in Raman spectra \cite{filik}, and their interaction
with an atomic force microscope tip \cite{gpzhang}. Diamondoid
functionalization \cite{mcintosh} and doping \cite{barnard} have also
been explored. In this investigation, the properties of functionalized
adamantane molecules were computed using the
``Vienna {\it ab initio} simulation package'' (VASP) \cite{kresse1}.  The
electronic exchange correlation potential was described with the density
functional theory/generalized gradient approximation (DFT/GGA) \cite{pbe}.
The electronic wave-functions were described by a Projector Augmented Wave
(PAW) \cite{kresse2} and expanded in a plane-wave basis set, with the kinetic
energy cutoff of 450 eV. Convergence in total energy was achieved when
it differed by less than 0.1 meV/atom between two self-consistent iterations.
The optimization of each configuration was performed until forces were
lower than 3 meV/\AA.

The isolated adamantane molecules, in pure or functionalized forms, were
placed in a cubic simulation cell with a fixed parameter of 17 \AA\ and
periodic boundary conditions. Such a large simulation cell guaranteed
negligible interactions between the molecule and its neighboring images. The
Brillouin zone was sampled by the $\Gamma$-point.
The enthalpy of formation ($\rm \Delta_{f} H$) of each adamantane-like
molecule was computed by:
\begin{equation}
\rm \Delta_{f}H (X_nC_{10-n}H_{16-n})=
E_{tot}(X_nC_{10-n}H_{16-n}) - nE(X) - (10-n)E(C) - (16-n)E(H)
\label{eform}
\end{equation}
where $\rm {E_{tot}(X_nC_{10-n}H_{16-n})}$ is the total energy of the
molecule, with n atoms of type X ($\rm X = B,N$), $\rm (10-n)$ carbon,
and $\rm (16-n)$ hydrogen atoms. The E(X), E(C), and E(H) are
the total energies, per atom, of respectively X, carbon, and hydrogen
elements, in their standard states. Those energies, computed within the
same methodology described in the previous paragraphs, were obtained from
the total energy of carbon in a diamond lattice, nitrogen in an isolated
N$_2$ molecule, hydrogen in an isolated H$_2$ molecule, and boron in a
trigonal crystalline structure. This procedure has been used in several
other systems to compute energies of formation \cite{assali,rolando}.

The crystal, in a zincblende
structure with a basis formed by tetra-bora-adamantane  and
tetra-aza-adamantane molecules, was simulated using the same theoretical
approximations and convergence criteria described earlier for the isolated
molecules. The Brillouin zone was sampled by a $8\times 8\times 8$ k-point grid.

\section{Pure and functionalized adamantane}
\label{sec3}

\begin{figure}[h]
\centering
\includegraphics[width=13cm]{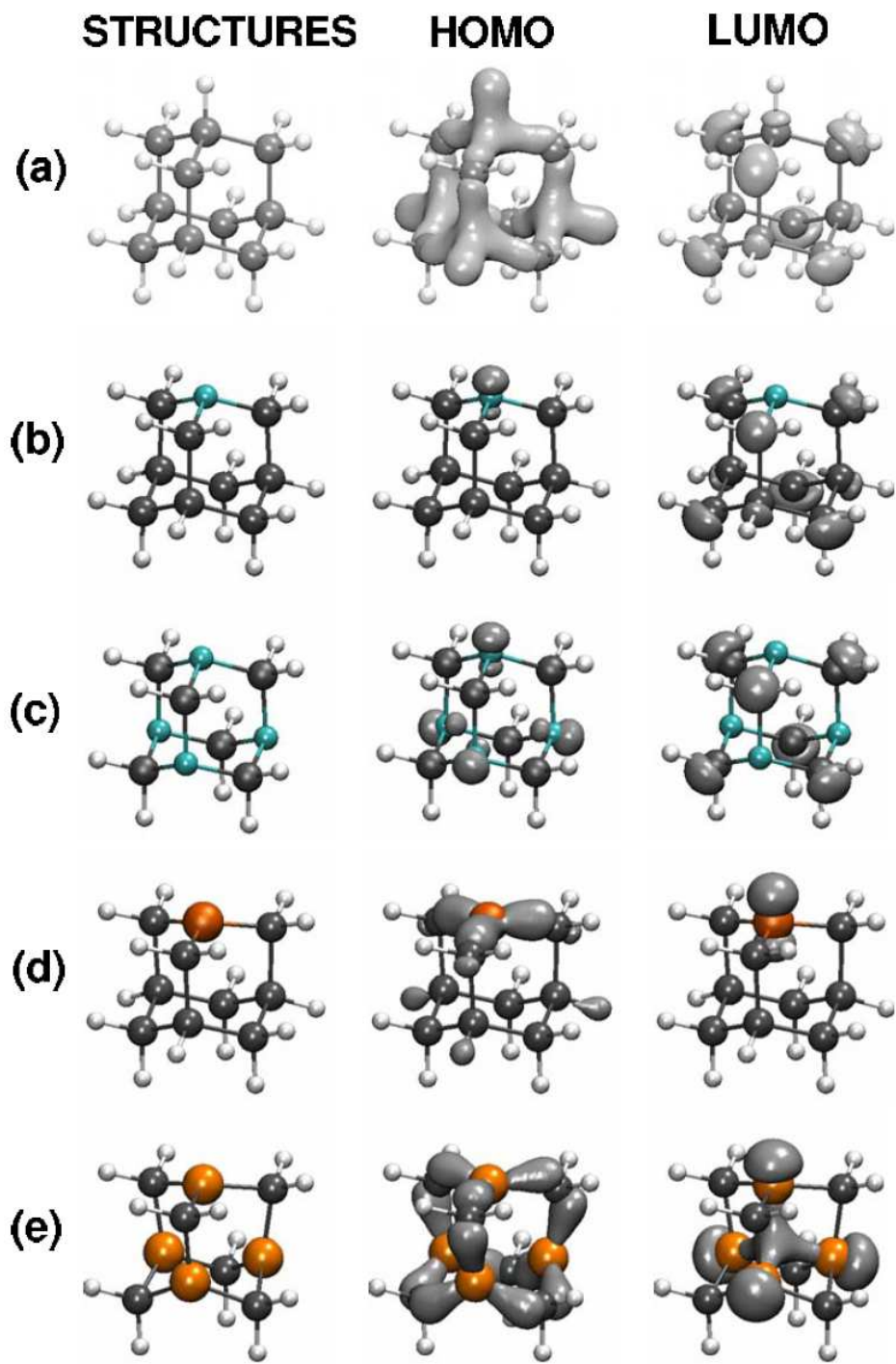}
\caption{(Color online) Optimized molecular structures
and probability density isosurfaces of HOMO and LUMO in:
(a) adamantane, (b) aza-adamantane, (c) tetra-aza-adamantane,
(d) bora-adamantane, and (e) tetra-bora-adamantane.
Black, light gray (blue), dark gray (orange),
and white spheres represent
carbon, nitrogen, boron, and hydrogen atoms, respectively.
Each isosurface corresponds to 30\% of the respective
maximum probability.}
\label{fig1}
\end{figure}

We discuss the properties of adamantane and the resulting changes
of incorporation single and multiple boron or nitrogen atoms.
There are two types of carbon atoms in adamantane, labeled C(1) and C(2).
A C(1) is bound to three C(2)'s and one H atom, while a C(2) is bound
to two C(1)'s and two H's. According to table \ref{tab1}, interatomic
distances and bond angles for all molecules investigated here were
in excellent agreement with available experimental data
\cite{vijay,kamp,mansson,bubnov}. In adamantane, the C(1)-C(2)
interatomic distances are 1.538 \AA, close to the respective value
in crystalline diamond (1.545 \AA). The C(1)-H and C(2)-H interatomic
distances are 1.105 \AA\ in average, close to typical C-H distances in
organic molecules. The average C-C-C bond angles are 109.5$^{\circ}$,
indicating the sp$^3$ character of carbon bonding. The H-C(2)-H bond
angles are 106.9$^{\circ}$, in agreement with another theoretical
investigation (106.8$^{\circ}$) \cite{gpzhang}. The electronic gap,
the difference in energy between the highest occupied molecular orbital
(HOMO) and the lowest unoccupied molecular orbital (LUMO),
is 5.7 eV, larger than the bandgap of crystalline diamond of 4.1 eV,
computed using the same methodology. The relative difference of
39\% between these two gap values is consistent with that obtained by
another theoretical investigation \cite{mcintosh}. Figure \ref{fig1}a
presents the relaxed structures and the respective HOMO and LUMO
probability density distributions. The HOMO is related to the C(1)-C(2)
plus the C(1)-H bonds, which is fully consistent with the one in hexamantane
\cite{mcintosh}. The LUMO is related to the C(1)-C(2) bonds with an
anti-bonding character, distributed in the back-bond of C(2) atoms.
Enthalpy of formation is -133.2 kcal/mol, demonstrating the strong
stability of adamantane.

Boron and nitrogen atoms are potential candidates to react with adamantane
molecules and lead to stable FBBs. They have essentially the same atomic
size of carbon, and therefore, they would cause small perturbations when
incorporated in carbon sites. Functionalization in the C(1) or C(2)
would provide respectively four or six active sites for intermolecular
bonding. Although, a larger number of active sites would be desirable, since
it could lead to stiffer nanomaterials \cite{mbb}, there are strong
evidences that functionalization with those atoms
is favorable in the C(1) sites \cite{bubnov,kamp}.

Incorporation of one nitrogen atom, replacing a C(1)-H group, to form
aza-adamantane (NC$_{9}$H$_{15}$), leads to important changes on the
properties of the original adamantane molecule. The N-C(2) interatomic
distance is 1.472 \AA, which is similar to the respective one in
aza-diamantane, in which theoretical calculations found a
value of 1.448 \AA\ \cite{mcintosh}. The nitrogen atom affects very
weakly the other C-C and C-H bonds. Nitrogen incorporation changes
the electronic structure of adamantane by reducing the gap from 5.6 to
3.6 eV. Nitrogen has five valence electrons and forms a trigonal
configuration, sharing three electrons with its neighboring C(2) atoms,
and keeping two electrons in a non-bonding p$_z$ configuration. While
the HOMO is associated with this non-bonding orbital, the LUMO is
essentially equivalent to that of adamantane, as shown in figure
\ref{fig1}b. This molecule is very stable, with an enthalpy of formation
of only 24.4 kcal/mol higher than that of adamantane.

Functionalization with three additional nitrogen atoms, replacing the
remaining C(1)-H groups of aza-adamantane, forms the
tetra-aza-adamantane molecule (N$_{4}$C$_{6}$H$_{12}$). The N-C(2)
distances  are equivalent to that in aza-adamantane, at 1.476 \AA.
The C-H interatomic distances and angles are slightly affected by
the inclusion of additional nitrogen atoms. In tetra-aza-adamantane,
the LUMO has the same character of adamantane and aza-adamantane, while
the HOMO is a degenerated state associated with the four nitrogen
non-bonding orbitals. The energy cost to form tetra-aza-adamantane
from adamantane is also small,  since enthalpy of formation is only
84.8 kcal/mol higher than that of adamantane. This value is in excellent
agreement with experimental data of 80.6 kcal/mol \cite{mansson}.

Replacing a C(1)-H group by one boron atom in adamantane forms
bora-adamantane (BC$_{9}$H$_{15}$). Although the C-C and C-H distances are
slightly affected by boron incorporation, the tetrahedral configuration of
the original C(1) site is deformed, leading to a near planar configuration
with a trigonal symmetry, with C(2)-B-C(2) and B-C(2)-C(1) bond angles at
116.6$^{\circ}$ and 98.1$^{\circ}$, respectively. B-C(2) interatomic distances
are 1.568 \AA, in agreement with other theoretical calculations for
bora-diamantane
at 1.555 \AA\ \cite{mcintosh}. Boron incorporation also reduces the gap of
adamantane. The charge distribution of HOMO is in the interstitial region
between B-C(2) bonds, while the LUMO is associated with the boron
non-bonding orbital, as shown in figure \ref{fig1}d.  Enthalpy of formation
of this molecule is only 19.1 kcal/mol higher than that of adamantane.
Incorporation of three boron atoms in bora-adamantane, replacing the
remaining C(1)-H groups, forms tetra-bora-adamantane
(B$_{4}$C$_{6}$H$_{12}$). HOMO and LUMO of this molecule are consistent with
the ones of bora-adamantane. The enthalpy of formation of
tetra-bora-adamantane is 37.5 kcal/mol higher than that of adamantane,
suggesting a strong stability for this molecule, although it has not
been synthesized so far.

We also simulated the bora-adamantane-aza-adamantane super-molecule
(H$_{15}$C$_{9}$B-NC$_{9}$H$_{15}$), which comprises an aza-adamantane
plus a bora-adamantane with  a B-N intermolecular bond. This B-N bond
distance is 1.737 \AA, in agreement with experimental value of
1.690 \AA\ \cite{bubnov}. All other interatomic distances and bond
angles (table \ref{tab1}) are in excellent agreement with experimental
data. Binding energy is 0.96 eV, showing that bonding between two
functionalized adamantane molecules is much stronger than that between
two pure molecules, around 0.1-0.2 eV \cite{mcintosh}.

\section{Functionalized adamantane molecular crystal}
\label{sec4}

Self-assembly of complex structures using nanoparticles has received
considerable attention in the literature \cite{mbb,iacovella}. Finding
appropriate BBs to allow precise and fast self-assembly still represents
a major challenge. In order to lead to stable nanostructures,
the BBs should carry a number of desired properties. The resulting
nanostructures should be stiff and rigid, therefore intermolecular bonding
between neighboring BBs should be strong, in addition to strong
intramolecular bonding. This rules out pure adamantane as a
potential FBB, since intermolecular bonding is generally weak, and although
a crystalline structure could be attained, it would not be enough
stiff \cite{sasagawa1}. On the other hand, functionalized adamantane could
satisfy both conditions. The strong intermolecular bonding could be attained
between the boron active sites in tetra-bora-adamantane and the nitrogen
ones in tetra-aza-adamantane, equivalent to bonding observed in the
aza-adamantane-bora-adamantane super-molecule. If those functionalized
molecules were
dissolved in solution, ionic interactions between the nitrogen and boron
sites in different molecules could favor self-assembly with great positional
precision. The resulting B$\leftarrow$N dative bonding would likely
guarantee stability and rigidity of the nanostructure up to room
temperature. Larger diamondoids could also be functionalized with boron
and nitrogen atoms, leading to additional building block types,
allowing more freedom in building complex structures.

\begin{figure}[h]
\centering
\includegraphics[width=11cm, angle=-90.0]{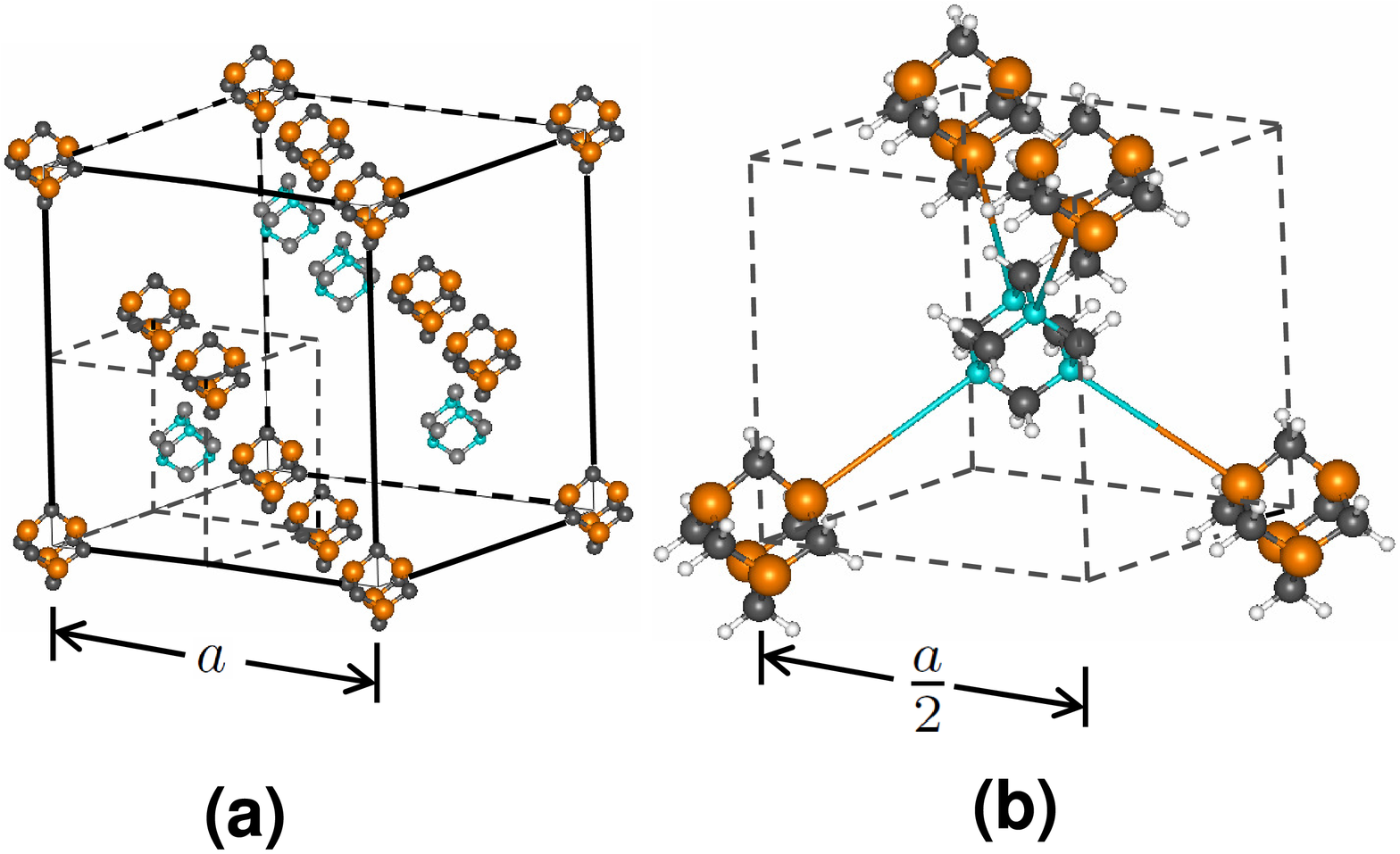}
\caption{(Color online)
Schematic representation of: (a) the zincblende structure of
the molecular crystal, composed of a tetra-bora-adamantane in the origin
plus a tetra-aza-adamantane in the (1/4,1/4,1/4)$a$ position,
where $a$ is the lattice parameter; (b) the tetrahedral
intermolecular B-N bonds.
For clarity, hydrogen atoms were removed in (a). (Coloring of
spheres is consistent with that in figure \ref{fig1}.)}
\label{fig2}
\end{figure}

The tetra-bora- and tetra-aza-adamantane isolated molecules
have tetrahedral symmetry and four chemically active sites, which allowed
to envision a hypothetical molecular crystal in a zincblende structure
(F$\bar{4}$3m space group), with those two molecules forming the basis.
Therefore, the crystal would be formed by a tetra-bora-adamantane sitting
in the lattice origin with a tetra-aza-adamantane sitting in the
(1/4,1/4,1/4)$a$ position, where $a$ is the lattice parameter.
Figure \ref{fig2} presents a schematic representation of this molecular
crystal. This crystal should be very stable, since boron and nitrogen atoms
would lie in a near tetrahedral environment and any boron (nitrogen) atom
in tetra-bora(tetra-aza)-adamantane would bind to one nitrogen (boron)
atom in a neighboring tetra-aza(tetra-bora)-adamantane molecule.
The intermolecular bonding in the crystal is essentially due to B-N
interactions.

\begin{figure}[h]
\centering
\includegraphics[width=12cm,angle=-90]{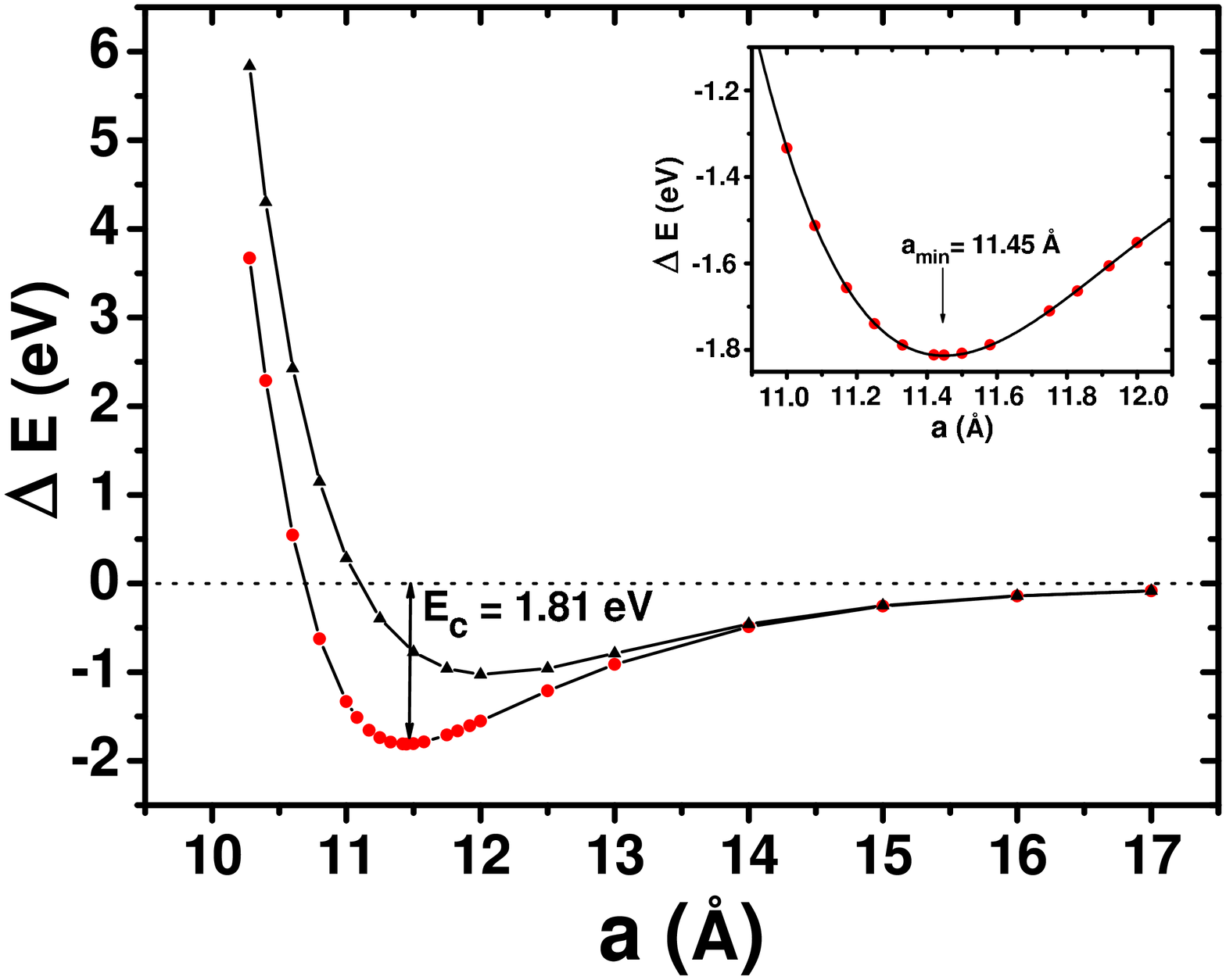}
\caption{Total energy variation ($\Delta$E) of the
tetra-aza-adamantane plus tetra-bora-adamantane
crystal, with respect to infinitely separated molecules,
as a function of the lattice parameter $a$,
with (dots) and without (triangles) relaxation of the molecular
internal degrees of freedom. The inset shows the energy
close to the minimum for the crystal with full relaxation.}
\label{fig3}
\end{figure}

Figure \ref{fig3} presents the crystal total energy as a function of the lattice
parameter. The reference energy is defined for infinitely separated
molecules, such that crystal formation leads to a large energy gain. We
initially simulated the system by varying the lattice parameter, constraining
the molecular internal degrees of freedom, such that only the B-N
distances varied. This already allowed a large cohesive energy
E$_{\rm c}$ = 1.02 eV per primitive cell at $a$ = 12.0 \AA. By releasing that
constrain, there was a further energy gain, and E$_{\rm c}$ = 1.81 eV at
$a$ = 11.45 \AA. The corresponding B-N distance is 1.907 \AA, which is
longer than that in bora-adamantane-aza-adamantane super-molecule,
discussed in previous section. This results from the competition among
the four B$\leftarrow$N bonds in each molecule. The internal configurations
of tetra-bora-adamantane and tetra-aza-adamantane molecules in the crystal
differed only slightly from those of isolated configurations. The
relevant relaxations were on the tetra-bora-adamantane molecule, where the
B-C bond distances and B-C-B bond angles changed from 1.589 \AA\  and
90.1$^{\circ}$  in the isolated molecule to 1.624 \AA\  and 102.5$^{\circ}$
in the crystal. In the crystal, boron relaxed toward a near tetrahedral
configuration by interacting with neighboring nitrogen atoms.

The stiffness of the resulting crystal could be determined by its
bulk modulus. We found a bulk modulus of 20 GPa, which is considerably
smaller than that of typical covalent solids, that ranged from 100 GPa,
in silicon, to around 400 GPa, in diamond and boron nitride. However,
it is still larger than the values for other molecular crystals \cite{day}.
Considering the small relaxation on intramolecular bonds in the molecular
crystal, the materials stiffness is essentially controlled by the variations
in the intermolecular B-N bonds. This is confirmed by comparing the density
of B-N bonds in c-BN with that in the molecular crystal. This density  is about
30 times smaller in the molecular crystal than in c-BN, being consistent
with a factor of about 20 between the respective bulk moduli.

\begin{figure}[h]
\centering
\includegraphics[width=15cm]{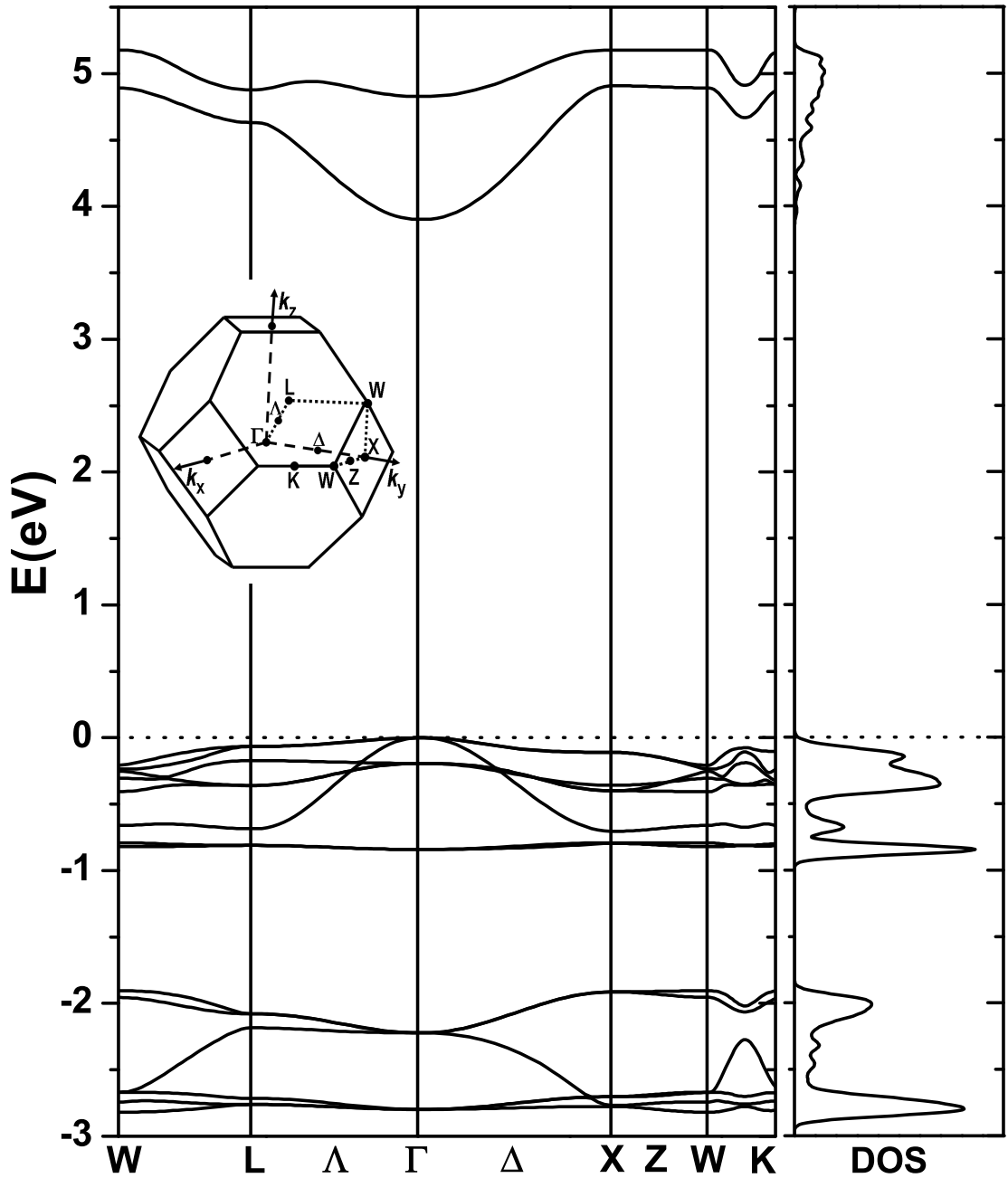}
\caption{Energy band structure along
several high symmetry directions and density of states (DOS) of the
molecular crystal schematically represented
in figure \ref{fig2}. The inset shows the first Brillouin zone
and the respective high symmetry points.}
\label{fig4}
\end{figure}

\begin{figure}[h]
\centering
\includegraphics[width=13.2cm, angle=0.0]{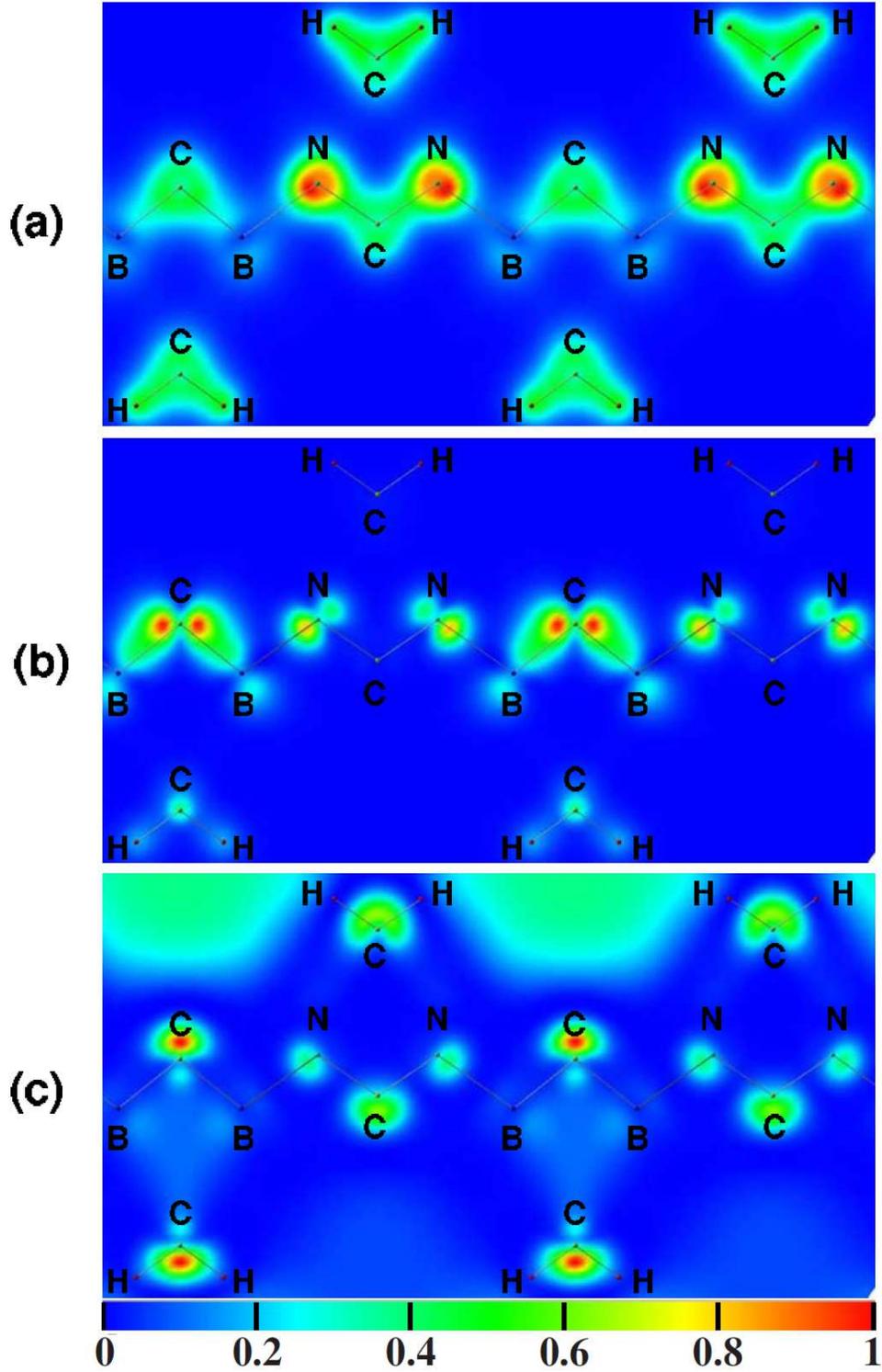}
\caption{(Color online)
Density probability isosurfaces in the (110) plane for the
molecular crystal: (a) total charge, (b) top of the valence band,
and (c) bottom of the conduction band. The results are represented
with respect to the maximum density probability
($\rho_{max}$ in e/\AA$^{3}$): (a) $\rho_{max} = 4.13$,
(b) $\rho_{max} = 0.54$, and (c) $\rho_{max} = 0.06$.}
\label{fig5}
\end{figure}

The band structure of the molecular crystal, depicted in Figure \ref{fig4},
shows a direct bandgap of 3.9 eV, which could be compared to 4.8 eV for
the direct gap of a pure adamantane molecular crystal \cite{sasagawa1}.
Both values should be considered as lower limits, since the DFT/GGA
framework generally underestimates gap energies. Corrections by those
authors \cite{sasagawa1} led to a 44\% increase in the gap, suggesting
corrections to the gap of this molecular crystal could lead to a value
around 5-6 eV. Figure \ref{fig5} shows the probability charge distribution
of the molecular crystal in the (110) plane. The top of the valence band
is described as a combination of the HOMO orbitals from isolated molecules:
the nitrogen non-bonding 2p orbital in tetra-aza-adamantane, that now is
associated with the B$\leftarrow$N dative bond, and the C-B orbital in the
tetra-bora-adamantane units. The bottom of the conduction band is associated
with the carbon atoms in the tetra-bora-adamantane molecules.

\section{Summary}
\label{sec5}

In summary, functionalized adamantane molecules have been investigated as
potential fundamental building blocks for nanostructure self-assembly.
Considering the enthalpies of formation, we
found that boron or nitrogen incorporation in adamantane molecules with
one (aza- or bora-adamantane) and four functional groups
(tetra-aza- or tetra-bora-adamantane) are very stable, although the last
one has not been synthesized so far. A hypothetical molecular crystal, formed
by  tetra-bora-adamantane plus tetra-aza-adamantane molecules,
in a zincblende structure, presented a reasonably large cohesive energy of
1.81 eV/primitive cell and bulk modulus of 20 GPa. These values are
considerably larger than those in typical molecular crystals, and may
provide stability and stiffness at room temperature. Any
defect in this hypothetical crystal, as result of imperfect self-assembly,
would likely cause only minor changes in the materials stiffness \cite{dumitrica}.
The electronic band structure of this crystal presented a direct and wide
bandgap of 3.9 eV, suggesting potential applications in opto-electronics.

\vspace{0.5cm}

\textbf{Acknowledgments}

The authors acknowledge support from the Brazilian Agency CNPq. The
calculations were performed at the computational facilities of
CENAPAD-S\~ao Paulo.


\newpage

\begin{table}[ht]
\caption{Properties of adamantane in pure
(C$_{10}$H$_{16}$) and functionalized forms (BC$_{9}$H$_{15}$,
B$_4$C$_{6}$H$_{12}$, NC$_{9}$H$_{15}$, N$_4$C$_{6}$H$_{12}$, and
H$_{15}$C$_{9}$B-NC$_{9}$H$_{15}$). The table presents the average bond angles (A)
and distances (d) between C(1)-C(2) and C(2)-X atoms,
where X is C(1), B or N in pure, boron or nitrogen functionalized
molecules, respectively. $\rm \triangle E_{H-L}$ is the HOMO-LUMO energy difference
and  $\rm \triangle_f H $ is the
enthalpy of formation, with adamantane as the
reference.  Distances, angles, energies, and enthalpies of
formation are given in \AA, degrees, eV, and kcal/mol, respectively.
Experimental data are given in parenthesis.}
\begin{center}
\begin{tabular}{cccccccc}
\hline \hline
    &~~{B$_{4}$C$_{6}$H$_{12}$}~~&~~{BC$_{9}$H$_{15}$}~~&
~~{C$_{10}$H$_{16}$}~~ & ~~{NC$_{9}$H$_{15}$}~~ & ~~{N$_{4}$C$_{6}$H$_{12}$}~~ &
\multicolumn{2}{c} {H$_{15}$C$_{9}$B-NC$_{9}$H$_{15}$}\\
                     & X = B & X = B& X = C(1)& X = N & X = N & X = B~~ & ~~X = N
                     \\
\hline
d[C(1)-C(2)] &  --  & 1.552  &  1.538  &  1.539  & --  &
1.543 & 1.537 \\
                               &      &  & (1.53)$^{(a)}$ &
                               &  & (1.536)$^{(d)}$ &  (1.534)$^{(d)}$ \\
d[C(2)-X]    &1.589 & 1.568  &  --     &  1.472  & 1.476
& 1.634 & 1.504 \\
                               &      & &            &  &
(1.473)$^{(b)}$  & (1.626)$^{(d)}$ & (1.503)$^{(d)}$ \\
A[C-C-C] &    -- & 110.6 & 109.5  & 108.8  & --
&  110.1 & 109.2 \\
                                   &       & &(109.45)$^{(a)}$ &
                                   &  & (110.1)$^{(d)}$ & (109.6)$^{(d)}$ \\
A[C(2)-X-C(2)]   & 117.2 & 116.6 & --    & 109.5  & 107.8 &
108.8 &  108.1  \\
                                   &       & &        &
                                    & (108.0)$^{(b)}$ &
                                   (108.6)$^{(d)}$ & (108.0)$^{(d)}$ \\
A[X-C(2)-C(1)]   &  --   & 98.1  & --      & 111.6  & --
& 107.5 &  112.1  \\
                                   &       &   &          &
                                    &  & (107.8)$^{(d)}$ & (112.5)$^{(d)}$ \\
A[X-C(2)-X]      &  90.1 & --    & --     & --     & 112.8
& -- & -- \\
                                   &        &      &        &        &
                                   (112.6)$^{(b)}$  & & \\
A[H-C(2)-H]      & 112.0 & 106.6 & 106.9  & 107.1  & 108.4
& 106.5 & 107.5 \\
                                   &       &       &        &        &
                                   (110.7)$^{(b)}$ & & \\
$\rm \triangle E_{H-L}$             & 2.3   & 4.8   & 5.7     & 3.7     & 4.4
& \multicolumn{2}{c} {4.1} \\
$\rm \triangle_f H $               & 37.5  & 19.1   & 0    & 24.4    & 84.8
& \multicolumn{2}{c} {21.2} \\
                                   &       &        &      &         &
                                   (80.6)$^{(c)}$  & & \\
\hline \hline
\end{tabular}
\end{center}
$^{(a)}$ Reference \cite{vijay}, $^{(b)}$  Reference \cite{kamp},
$^{(c)}$  Reference \cite{mansson}, $^{(d)}$  Reference \cite{bubnov}.
\label{tab1}
\end{table}

\pagebreak

\pagebreak

\pagebreak

\pagebreak

\pagebreak

\end{document}